\documentclass[twocolumn,superscriptaddress,amsmath,ammsymb]{revtex4-2}
\usepackage{epsfig}
\usepackage{amsmath}
\usepackage{amssymb}
\usepackage{graphicx}
\usepackage{dcolumn}
\usepackage{bm}
\usepackage[usenames]{color}
\usepackage[breaklinks]{hyperref}

\hypersetup{colorlinks=true,allcolors=blue}

\begin{document}
	
\title{Magneto-optic signatures in the gapped Dirac semimetal with mixed linear and parabolic dispersions of ZrTe$_5$}

\author{Le You} 
\affiliation{School of Science, Jiangnan University, Wuxi 214122, China}

\author{Zi-Yue Zhang} 
\affiliation{School of Science, Jiangnan University, Wuxi 214122, China}

\author{Yi-Xiang Wang}
\email{wangyixiang@jiangnan.edu.cn}
\affiliation{School of Science, Jiangnan University, Wuxi 214122, China}
\affiliation{School of Physics and Electronics, Hunan University, Changsha 410082, China}

\date{\today}	
	
\begin{abstract}
In this paper, we give a systematic theoretical study on the Landau levels (LLs) and magneto-optical conductivity Re$(\sigma_{\alpha\alpha})$ in a gapped Dirac semimetal model with mixed linear and parabolic dispersions under a magnetic field, which was recently proposed by Jiang \textit{et al.} [Phys. Rev. Lett. {\bf125}, 046403 (2020)] to explain the experimental magnetoinfrared spectroscopy in the three-dimensional ZrTe$_5$ crystal.  We find that the strong magnetic field can drive the LLs become noninverted and thus the strong topological insulator phase in ZrTe$_5$ turns to be a trivial insulator.  In the different magnetic field regions, the density of states and Re$(\sigma_{\alpha\alpha})$ can exhibit distinct signatures.  Moreover, when the magnetic field is weak, a qualitative relation in Re$(\sigma_{zz})$ between the peaks at the saddle points is revealed as Re$(\sigma_{zz}^{\zeta_n})>$Re$(\sigma_{zz}^\Gamma)$, which is in good agreement with the experiment.   
\end{abstract} 

\maketitle

\section{Introduction} 

The topological insulators (TIs) that are characterized by the bulk band inversions have aroused the ongoing interests in the past decades~\cite{M.Z.Hasan, X.L.Qi, O.Vafek, N.P.Armitage}.  Among the search of various topological materials~\cite{B.Q.Lv}, the three-dimensional (3D) ZrTe$_5$, a material of weakly interlayer bonding, owns the unconventional topological properties and has therefore ignited renewed interests.  Actually, the 3D ZrTe$_5$ has been studied since 1980s due to its large thermoelectric power and resistivity anomaly~\cite{S.Okada, E.F.Skelton, T.E.Jones, T.M.Tritt}.  In 2014, with the \textit{ab initio} calculations, Weng \textit{et al.} predicted that a single-layer ZrTe$_5$ was a candidate for the large-gap quantum spin Hall insulator~\cite{H.M.Weng}, and further argued that its 3D bulk crystal was located near the phase boundary between a weak and a strong topological insulator~\cite{H.M.Weng}. 

On the other hand, the experimental verifications of the low-energy property in 3D ZrTe$_5$ crystal remain highly controversial. 
The (magneto-) optical spectroscopy experiments suggested that ZrTe$_5$ might be a Dirac semimetal, due to its linear dependence of the optical conductivity on the photon frequency~\cite{R.Y.Chen2015a, R.Y.Chen2015b, Z.G.Chen, B.Xu, Y.Jiang2017}. 
By using the scanning tunneling microscopy or spectroscopy (STM/STS) and angle-resolved photoemission spectroscopy (ARPES), a large bulk gap in ZrTe$_5$ was identified, together with the topological edge states at the surface edge, showing that ZrTe$_5$ was a weak 3D TI~\cite{R.Wu, X.B.Li, H.Xiong}.  
Another ARPES and STM study reported that ZrTe$_5$ was a strong TI~\cite{G.Manzoni}, where both the bulk and surface states were identified at the top of the valence band.  The complication in the topological character of ZrTe$_5$ may be attributed to its interlayer distance~\cite{H.M.Weng}, temperature~\cite{B.Xu}, purity and other factors. 
In contrast, the chiral magnetic effect and nontrivial Berry phase were clearly observed in ZrTe$_5$ through the magnetotransport measurements~\cite{Q.Li, G.L.Zheng, J.L.Zhang}. 

Although the early studies debated about the ground state property in ZrTe$_5$, a consensus was reached that the Dirac cone at the $\Gamma$ point of the Brillouin zone was 3D.  However, several experimental studies~\cite{J.L.Zhang, E.Martino, Y.Jiang2020} reported that there exist discrepancies between the observations and the 3D Dirac model predictions~\cite{P.Hosur, A.Bacsi, Ashby, C.J.Tabert}, and thus began to question the dimensionality of the Dirac cone in ZrTe$_5$.  In Ref.~\cite{J.L.Zhang} of the magnetotransport measurement under hydrostatic pressure and in Ref.~\cite{E.Martino} of the optical spectroscopy experiment, the Dirac cone was suggested to be not 3D, but only two-dimensional (2D).  Recently, Jiang \textit{et al.}~\cite{Y.Jiang2020} performed a magnetoinfrared spectroscopy measurement at liquid helium temperature 4.2 K in ZrTe$_5$, where an additional set of optical transitions from a band gap next to the $\Gamma$ point was observed, and thus proposed a gapped Dirac semimetal model with mixed linear and parabolic dispersions to describe its ground state.  They concluded that the bulk bands were inverted and hence ZrTe$_5$ was a strong TI.  However, a comprehensive theoretical analysis of the magneto-optical conductivity of the proposed model is still lacking, which motivates the present work. 

In this paper, we make a systematic study on the gapped Dirac semimetal model under a magnetic field as well as on how to characterize the bulk states from the magneto-optical signatures.  We first analyze the evolutions of the Landau levels (LLs) and the density of states (DOS) with the magnetic field.  Then by using the Kubo formula, we calculate the magneto-optical conductivity Re$(\sigma_{\alpha\alpha})$ of the system, with the impurity scatterings being included phenomenologically.  Our main results are as follows:   
(i) The evolutions of the LL structures are caused by the opposite changes of the two characteristic quantities $\Delta_n$ and $\Delta_{ns}$ [Eqs.~(\ref{quantity1}) and~(\ref{quantity2})] with the magnetic field.  At strong magnetic field, the LLs are noninverted and thus the strong TI becomes a trivial insulator. 
(ii) In the different magnetic field regions, the DOS and Re$(\sigma_{\alpha\alpha})$ can exhibit distinct signatures. 
(iii) When the magnetic field is weak, we find that at the $\Gamma$ and $\zeta_n$ points, for Re$(\sigma_{zz})$, there exist a qualitative relation between the two peaks, Re$(\sigma_{zz}^{\zeta_n})>$Re$(\sigma_{zz}^\Gamma)$, which is consistent with the experiment~\cite{Y.Jiang2020}, whereas for Re$(\sigma_{xx})$, the two peaks have almost equal heights, Re$(\sigma_{xx}^{\zeta_n})\simeq$Re$(\sigma_{xx}^\Gamma)$.  These conclusions also hold in the intermediate magnetic field region and behave as the intrinsic properties of the gapped Dirac semimetal model.  Our work demonstrate that the magnetic field can provide an effective means to modulate the physical property of the model, which can thus help understand the topological ground state in 3D ZrTe$_5$ crystal.

\section{LLs and DOS}

We start from the gapped Dirac semimetal model with mixed linear and parabolic dispersions.  In the four-component basis $\begin{pmatrix}
|+,\uparrow\rangle& 
|+,\downarrow\rangle& 
|-,\uparrow\rangle& 
|-,\downarrow\rangle
\end{pmatrix}^T$, the Hamiltonian is~\cite{Y.Jiang2020} ($\hbar=1$)
\begin{align}
H(\boldsymbol k)
=&v(k_x\tau_x\otimes\sigma_z+k_y\tau_y\otimes I )
+v_z k_z \tau_x\otimes\sigma_x 
\nonumber\\
&+[M-\xi(k_x^2+k_y^2)-\xi_zk_z^2]\tau_z\otimes I.
\label{Ham}
\end{align}
Here $\sigma$ and $\tau$ are the Pauli matrices acting on the spin and orbit degrees of freedom, respectively.  $v$ and $v_z$ are the Fermi velocities, $\xi$ and $\xi_z$ are the band inversion parameters, and $M$ denotes the Dirac mass.  Although the 3D ZrTe$_5$ is anisotropic, we assume the isotropy in the $x$-$y$ plane, $v_x=v_y=v$ and $\xi_x=\xi_y=\xi$.  In the following, we take the model parameters as $(v,v_z)=(6,0.5)\times10^5$ m/s, $(\xi,\xi_z)=(0.1,0.2)$ eV$\cdot$nm$^2$, and $M=7.5$ meV.  As $v_z\ll v$ and $\zeta_z\sim\zeta$, the 3D bands own a much slower dispersion in the $z$ direction than those in the $x-y$ plane.  

With no magnetic field,the Hamiltonian $H(\boldsymbol k)$ supports a pair of Dirac points at the $\zeta_\pm$ points, $(k_\parallel,\pm k_{zD})=(0,\pm\sqrt{\frac{M}{\zeta_z}-\frac{v_z^2}{2\xi_z^2}})$, as reported in our previous study~\cite{Y.X.Wang2021}.  It can be checked that the energy gaps at the $\Gamma$ and $\zeta$ points are given as $\Delta_\Gamma=2M=15$ meV and $\Delta_\zeta=2[(M-\xi_zk_{zD}^2)^2+v_z^2k_{zD}^2]^\frac{1}{2}=11.5$ meV, respectively, which agree with the experimental observations~\cite{Y.Jiang2020}.  Compared with the previously proposed 2D conical model to describe ZrTe$_5$~\cite{E.Martino,Y.X.Wang2020,Z.Rukelj}, where the linear terms occur in the $x-y$ plane and the parabolic term is along in the $z$ direction, $H(\boldsymbol k)$ in Eq.~(\ref{Ham}) incorporates the linear as well as the parabolic terms in all three directions.  When $v_z=\xi=0$, $H(\boldsymbol k)$ reduces to the 2D conical model. 

When a magnetic field $B\hat z=\nabla\times \boldsymbol A$ is incorporated in the system, we neglect the spin Zeeman effect and consider the orbital effect via the Peierls substitution, $\boldsymbol k\rightarrow \boldsymbol \pi=\boldsymbol k-e\boldsymbol A$ and $[\pi_x,\pi_y]=ieB$.  The ladder operators are defined as $a=\frac{l_B}{\sqrt{2}}(\pi_x+i\pi_y)$ and $a^\dagger=\frac{l_B}{\sqrt{2}}(\pi_x-i\pi_y)$, satisfying $a|n\rangle=\sqrt n|n-1\rangle$ and $a^\dagger|n\rangle=\sqrt{n+1}|n+1\rangle$, where $l_B=\sqrt{\frac{1}{eB}}=\frac{25.6}{\sqrt B}$nm is the magnetic length and $|n\rangle$ denotes the basis of the occupation number representation.  With the help of the unitary matrix 
\begin{align}
U=\begin{pmatrix}
1& 0& 0& 0\\
0& 0& 1& 0\\
0& 1& 0& 0\\
0& 0& 0& 1
\end{pmatrix}, 
\end{align}
the basis is transformed as $\begin{pmatrix}
|+,\uparrow\rangle& 
|-,\uparrow\rangle& 
|+,\downarrow\rangle& 
|-,\downarrow\rangle
\end{pmatrix}^T$ and the Hamiltonian becomes 
\begin{align}
H_B=UHU^\dagger=\begin{pmatrix}
L(k_z)& \frac{\sqrt2 v}{l_B} a^\dagger& 0& v_zk_z
\\
\frac{\sqrt2 v}{l_B} a& -L(k_z)& v_zk_z& 0
\\
0& v_zk_z& L(k_z)& -\frac{\sqrt2 v}{l_B} a
\\
v_zk_z& 0& -\frac{\sqrt2 v}{l_B} a^\dagger& -L(k_z) 
\end{pmatrix}, 
\label{HB}
\end{align}
where $L(k_z)=M-\frac{2\xi}{l_B^2}(a^\dagger a+\frac{1}{2})-\xi_zk_z^2$.  

To solve $H_B$, we use the trial wavefunction $\psi_n=\begin{pmatrix}c_n^1|n\rangle&
c_n^2|n-1\rangle&
c_n^3|n-1\rangle&
c_n^4|n\rangle
\end{pmatrix}$ and then $H_B$ is written as 
\begin{align}
H_{Bn}=&\begin{pmatrix}
L_n& b_n & 0& v_z k_z
\\
b_n& -L_n& v_z k_z& 0
\\
0& v_z k_z& L_n& -b_n
\\
v_z k_z& 0& -b_n& -L_n   
\end{pmatrix}+H_z, 
\label{HBn}  
\end{align}
with the Zeeman term
\begin{align}
H_z=-\frac{\xi}{l_B^2}\sigma_z\otimes I, 
\label{Zeeman}
\end{align} 
where $L_n=M-\frac{2n\xi}{l_B^2}-\xi_zk_z^2$, $b_n=\frac{\sqrt{2n} v}{l_B}$.  For $n\geq1$, the eigenenergy and wavefunction are solved as  
\begin{align}
\varepsilon_{n\lambda s}(k_z)
=\lambda\Big[\Big(\sqrt{L_n^2 + b_n^2}+s\frac{\xi}{l_B^2}\Big)^2
+v_z^2 k_z^2\Big]^\frac{1}{2}, 
\label{epsilonn}
\end{align}
and
\begin{align}
&\psi_{n\lambda s}=c_{n\lambda s}
\nonumber\\
&\times\begin{pmatrix}
-(\frac{\xi}{l_B^2}+s\sqrt{L_n^2+b_n^2}-\varepsilon_{n\lambda s})|n\rangle
\\
\big[b_n+\frac{1}{b_n}(L_n+s\sqrt{L_n^2+b_n^2})(\frac{\xi}{l_B^2}+L_n-\varepsilon_{n\lambda s})\big]|n-1\rangle
\\
\frac{1}{b_n} v_z k_z (-s\sqrt{L_n^2+b_n^2}-L_n)|n-1\rangle
\\
v_z k_z|n\rangle
\end{pmatrix}, 
\label{psin}
\end{align}
with $c_{n\lambda s}$ being the normalized coefficient, and the index $\lambda,s=\pm1$.  For $n=0$, we have 
\begin{align}
\varepsilon_{0\lambda}(k_z)
=s\Big[\big(M-\frac{\xi}{l_B^2}-\xi_z k_z^2\big)^2+v_z^2k_z^2\Big]^\frac{1}{2},   
\label{epsilon0}
\end{align}
and 
\begin{align}
\psi_{0\lambda}=c_{0\lambda}\begin{pmatrix}
\big[(M-\frac{\xi}{l_B^2}-\xi_zk_z^2)+\varepsilon_{0\lambda}\big]|0\rangle\\ 0\\ 0\\ k_z|0\rangle
\end{pmatrix}. 
\label{psi0}
\end{align}
We can see that at the $\Gamma$ point, $H_{Bn}$ in Eq.~(\ref{HBn}) is block diagonalized, so the upspin and downspin components become decoupled.  As a result, for $n\geq1$, when $\lambda s=1(-1)$, the wavefunction contains only the downspin (upspin) component and for $n=0$, when $\lambda=1(-1)$, the wavefunction contains only the upspin (downspin) component.  Importantly, the eigenenergy $\varepsilon_{n+s}(k_z=0)$ will increase with $B$ and move away from the zero energy, while $\varepsilon_{0+}(k_z=0)$ will decrease with $B$ and move close to the zero energy.  

With the help of the Green's function $G(k_z,z)=[z-H(k_z)]^{-1}$, the DOS of the system is defined as~\cite{Y.X.Wang2017}
\begin{align}
D(\varepsilon)&=-\frac{1}{\pi V}\sum_{\boldsymbol k} \text{Im}G(k_z,z=\varepsilon+i\tau^{-1})
\nonumber\\
&=-\frac{1}{4\pi^3 l_B^2}\int_\infty dk_z  
\text{Im}G(k_z,z=\varepsilon+i\tau^{-1}), 
\label{DOS}
\end{align}
where $\tau$ is the scattering rate and its inverse $\tau^{-1}$ phenomenologically represents the impurity scatterings-induced linewidth broadening.  The factor $\frac{1}{2\pi l_B^2}$ accounts for the LL degeneracy in the $x-y$ plane, thus the DOS increases with the magnetic field.  The cases are the same for the optical conductivity studied in the next section.  To obtain the DOS, one requires to integrate over $k_z$ in Eq.~(\ref{DOS}), which, however, are not readily accessible.  Instead, we resort to the numerical integrations.  Then the limiting clean case actually corresponds to a very low linewidth, such as $\tau^{-1}=0.001$ meV. 

\begin{figure}
	\includegraphics[width=8.8cm]{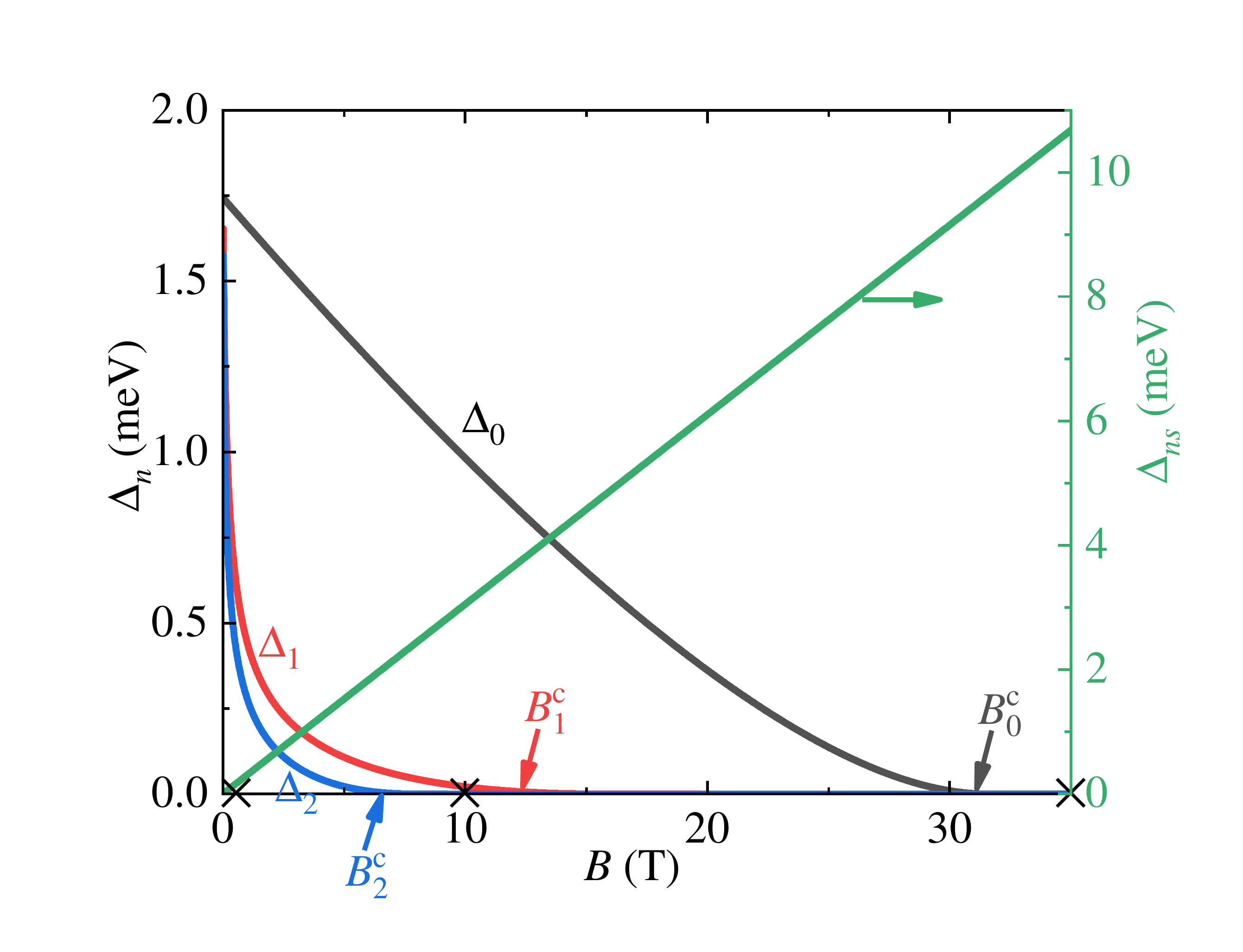}
	\caption{(Color online) $\Delta_{n=0,1,2}$ and $\Delta_{ns}$ as a function of the magnetic field $B$.  $B_n^c$ marks the critical magnetic field at which $\Delta_n=0$, meaning that the $\zeta_n$ point merges with the $\Gamma$ point of the $n-$LL.  The crosses label the values of the magnetic field used in the following calculations.  Note the double $y-$axis in the plot.}
	\label{Fig1}	
\end{figure}

\begin{figure}
	\includegraphics[width=9cm]{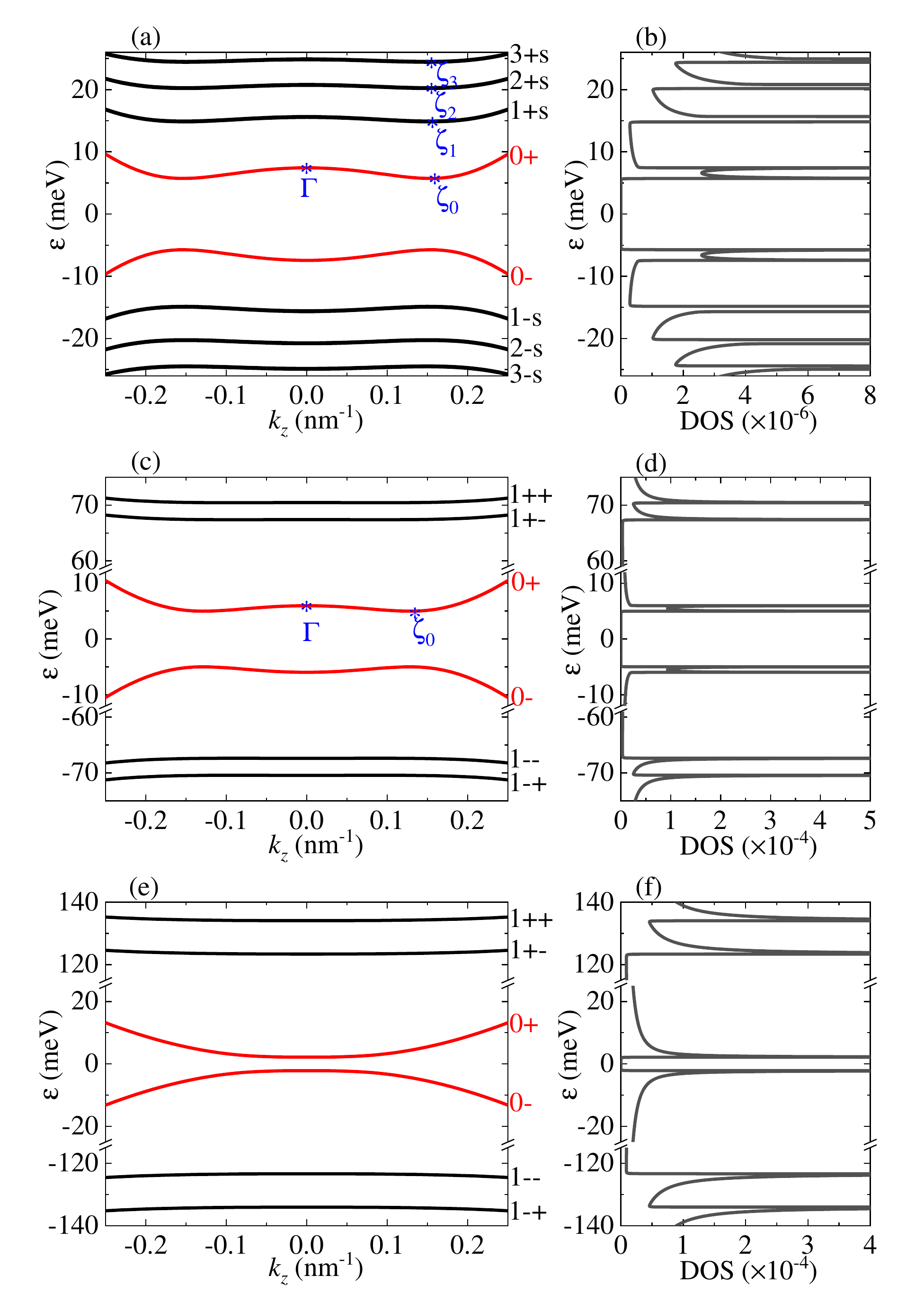}
	\caption{(Color online) The LL dispersions in (a), (c) and (e), and the corresponding DOS in (b), (d) and (f).  The LL index $n\lambda s$ are labeled in the dispersions, where in (a), the $s=\pm1$ branches are degenerated.  We set the magnetic field $B=0.4$ T in (a) and (b), $B=10$ T in (c) and (d), and $B=35$ T in (e) and (f). }
	\label{Fig2}
\end{figure}

We analyze the evolutions of the LLs and DOS with the magnetic field $B$.  To see the evolutions more clearly, we define two characteristic quantities, 
\begin{align}
&\Delta_n=\varepsilon_{n+-}(\Gamma)-\varepsilon_{n+-}(\zeta_n), 
\label{quantity1}
\\
&\Delta_{ns}=\varepsilon_{n++}(\Gamma)-\varepsilon_{n+-}(\Gamma)=\frac{2\xi}{l_B^2}\propto B,
\label{quantity2}
\end{align}
where $\zeta_n$ denotes the additional energy saddle point (extremum) of the $n$-th LL, other than the $\Gamma$ point, if exists [Figs.~\ref{Fig2}(a) and (c)].  The quantity $\Delta_n$ gives the energy difference between the $\Gamma$ and $\zeta_n$ points of the $(n+-)$ LL, and $\Delta_{ns}$ represents the energy difference between the $(n++)$ and $(n+-)$ LL branches at the $\Gamma$ point.  $\Delta_{ns}$ gives the Zeeman splitting between the two spin branches, which does not depend on the index $n$ but is proportional to $B$. 

When the magnetic field increases, the $\zeta_n$ point of the $n$-th LL will move to the $\Gamma$ point and finally merge with it.  So the energy splitting between the two saddle points diminishes and the linear components in the system become dominant.  In Fig.~\ref{Fig1}, we plot $\Delta_{n=0,1,2}$ and $\Delta_{ns}$ as a function of $B$, where
$\Delta_{0,1,2}$ decrease with $B$ and becomes vanishing at the critical $B_n^c$, while 
$\Delta_{ns}$ increases with $B$.  The critical magnetic fields satisfy $B_2^c<B_1^c<B_0^c$, meaning that with the increase of $B$, the merging of the $\zeta_n$ and $\Gamma$ points occurs successively for the LLs of decreasing $n$, and finally for the $n=0$ LL.  Besides through $\Delta_0=0$, the critical magnetic field $B_0^c$ can also be obtained by just setting the $k_z$ value at $\zeta_0$ of the 0+ LL to be zero, which gives
\begin{align}
B_0^c=\frac{1}{e\xi}\Big(M-\frac{v_z^2}{2\xi_z}\Big).  
\label{critB}
\end{align}
Eq.~(\ref{critB}) tells us that $B_0^c$ is closely related to the band inversion parameters; it decreases with $\xi$, but increases with $\xi_z$.  For the chosen parameters, we have $B_0^c=31.45$ T.  Due to the interplay of $\Delta_n$ and $\Delta_{ns}$, three regions can be separated for the magnetic field: (i) the weak field region, $B\ll B_0^c$, (ii) the intermediate field region, $B_2^c<B<B_0^c$ and (iii) the strong field region, $B>B_0^c$. 

In Fig.~\ref{Fig2}, we plot the LL dispersions and the corresponding DOS in the three regions, where the magnetic field is chosen as $B=0.4$ T in (a)$-$(b), $B=10$ T in (c)$-$(d), and $B=35$ T in (e)$-$(f), as labeled by the crosses in Fig.~\ref{Fig1}.  Experimentally, $B=0.4$ T is actually the magnitude implemented in the magnetoinfrared spectroscopy measurements~\cite{Y.Jiang2020}.  (i) When the magnetic field $B$ is weak [Fig.~\ref{Fig2}(a)], for $n\geq1$, the Zeeman splitting $\Delta_{ns}$ is close to zero, so the $s=\pm1$ LL branches are nearly degenerate.  By contrast, the energy difference $\Delta_n$ is large enough to be distinguished: the $n=0$ LL owns two distinctive energy extrema at the $\Gamma$ point and the $\zeta_0$ point of $k_{z0}=\pm\big(\frac{M}{\xi_z}-\frac{\xi}{\xi_z l_B^2}-\frac{v_z^2}{2\xi_z^2}\big)^\frac{1}{2}$, while the $n\geq1$ LLs also own two energy extrema at the $\Gamma$ point and the $\zeta_n$ point of $k_{zn}\simeq\pm\big(\frac{M}{\xi_z}-\frac{2n\xi}{\xi_z l_B^2}-\frac{v_z^2}{2\xi_z}\big)^{\frac{1}{2}}$.  We note that the $\zeta_n$ point is close to $\zeta_0$ under a weak $B$ [Fig.~\ref{Fig2}(a)].  The energy extrema of the LLs are related to the Van-Hove singularities of the DOS.  Thus the double-peak structures of each $n\leq2$ LLs are observed in the DOS [Fig.~\ref{Fig2}(b)], while those of the $n\geq3$ LLs cannot be distinguished due to the vanishing $\Delta_{n\geq3}$.  
(ii) At the intermediate $B$ [Fig.~\ref{Fig2}(c)], $\Delta_0\sim1$ meV but $\Delta_{1,2}$ are close to zero, so two energy extrema exist in the $n=0$ LL, but not in the $n\geq1$ LLs.  On the other hand, $\Delta_{ns}\sim3$ meV, meaning that the $s=\pm1$ branches are not degenerate anymore.  Therefore, the double-peak structure can be found in the DOS for the $n=0$ LL as well as $n\geq1$ LLs [Fig.~\ref{Fig2}(d)], where the former originates from the energy difference between the two saddle points and the latter is due to the Zeeman splitting.   
(iii) At the strong $B$ [Fig.~\ref{Fig2}(e)], for the $n=0$ LL, only one energy extremum occurs at the $\Gamma$ point.  This means that all LLs are noninverted and the system becomes a trivial insulator.  For the $n\geq1$ LLs, as $\Delta_{ns}$ increases, the $s=\pm1$ branches are further splitted.  Therefore in the DOS [Fig.~\ref{Fig2}(f)], the single-peak and double-peak structures can be found in the $n=0$ LL and $n\geq1$ LLs, respectively. 

According to the above analysis, when the magnetic field lies in the different regions, there may exist single-peak or double-peak structures in the DOS of the $n=0$ LL as well as $n\geq1$ LLs, which will have important significance in the magnetic-optical conductivity, as studied below.

\section{Magneto-optical conductivity} 

\begin{figure*}
	\centering
	\includegraphics[width=18.2cm]{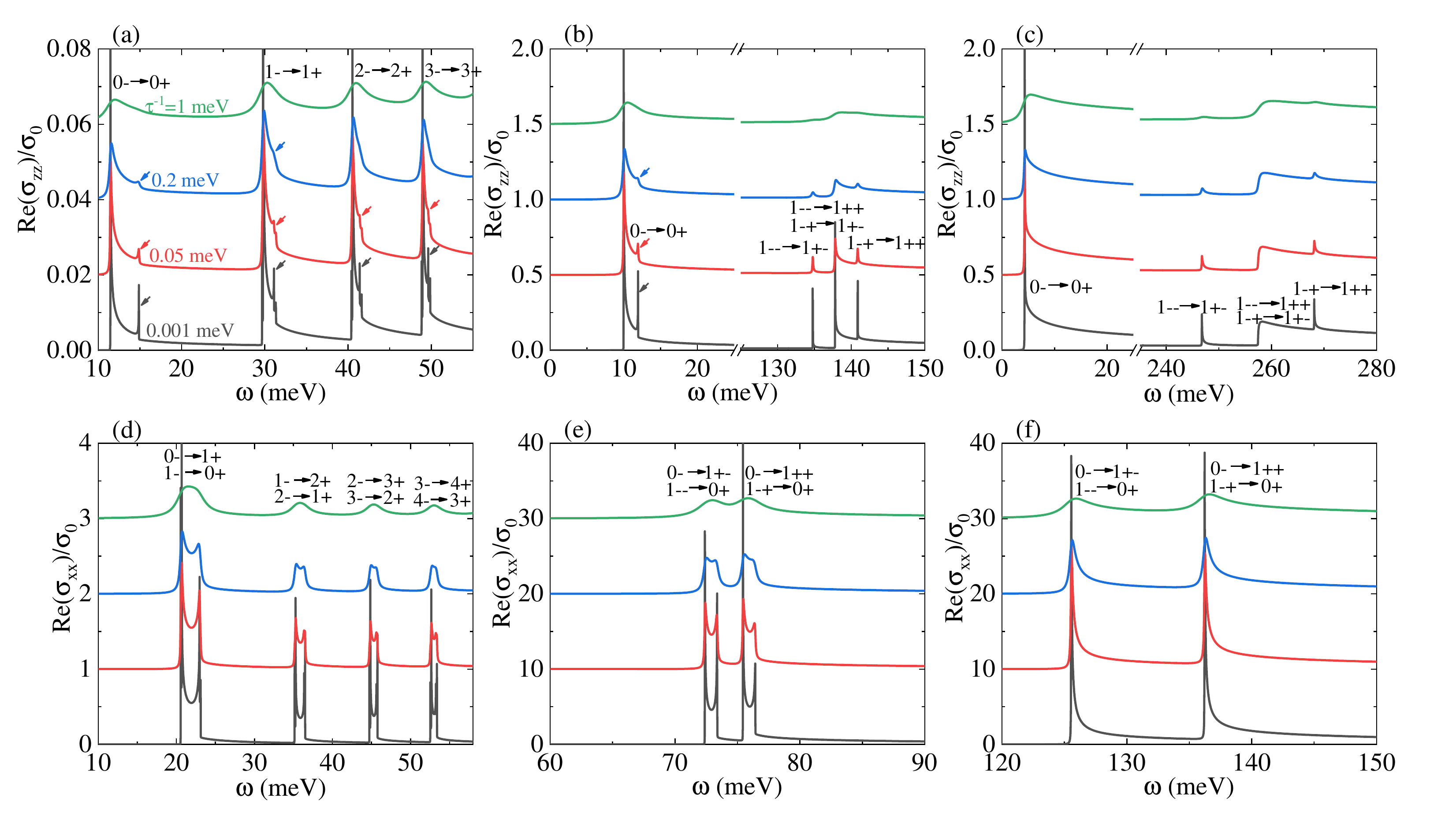}
	\caption{(Color online) The magneto-optical conductivity Re$(\sigma_{\alpha\alpha})$ (in unit of $\sigma_0=\frac{e^2}{h}$) vs the photon frequency $\omega$ for different linewidth $\tau^{-1}$, with $\alpha=z$ in (a)$-$(c) and $\alpha=x$ in (d)$-$(f).  We set the magnetic field $B=0.4$ T in (a) and (d), $B=10$ T in (b) and (e), and $B=35$ T in (c) and (f).  The LL transitions are labeled for the resonant  peaks.  In (a) and (b), the unambiguous peaks at the $\Gamma$ point are shown by the arrows, which are weaker than those at the $\zeta_n$ point.  For clarity, the lines are vertically shifted.  The legends are the same in all figures.}
	\label{Fig3}
\end{figure*}

The magneto-optical conductivity has been shown to be a valuable technique in identifying the band structures as well as the topological properties of various materials~\cite{Ashby, X.Lu, D.K.Mukherjee, W.Duan2019, W.Duan2020, S.Tchoumakov}.  We focus on the diagonal terms of the optical conductivity tensor, which can be calculated by using the linear-response Kubo formula~\cite{Y.X.Wang2017, Y.X.Wang2020}, 
\begin{align}
\sigma_{\alpha\alpha}(\omega)=&\frac{-i}{2\pi l_B^2}
\int_{-\infty}^\infty \frac{dk_z}{2\pi}
\sum_{n,n'}\sum_{\lambda,\lambda'}\sum_{s,s'}
\frac{f(\varepsilon_{n\lambda s})-f(\varepsilon_{n'\lambda's'})}
{\varepsilon_{n\lambda s}-\varepsilon_{n'\lambda's'}}
\nonumber\\
&\times
\frac{|\langle\psi_{n\lambda s}|J_\alpha |\psi_{n'\lambda's'}\rangle|^2}
{\omega+\varepsilon_{n\lambda s}-\varepsilon_{n'\lambda's'}+i\tau^{-1}}.  
\end{align}
Here $f(x)$ is the Fermi-Dirac distribution function, $J_\alpha=-ie[r_\alpha,H]$ is the current density operator and $\alpha=x,z$ denotes the direction that the optical field acts on.  In the calculations, we choose the system at half filling and zero temperature, so the index $\lambda=-1$ and $\lambda'=1$. 

The matrix elements $\langle\psi_{n\lambda s}|J_\alpha |\psi_{n'\lambda's'}\rangle$  of the current density operator along the $\alpha$ direction are evaluated with the eigenstates $\psi_{n\lambda s}$, whose nonvanishing value determines the selection rules for the LL transitions.  For a magnetic field along the $z$ direction, when the optical field acts in the parallel $z$ direction, $\boldsymbol E\parallel\boldsymbol B$, the matrix elements are 
\begin{widetext}
\begin{align}
\langle\psi_{n\lambda s}|J_z|\psi_{n'\lambda's'}\rangle
=&e\Big[
2\xi_z k_z  (-c_{n\lambda s}^1 c_{n\lambda's'}^1 
+c_{n\lambda s}^2 c_{n\lambda's'}^2 
-c_{n\lambda s}^3 c_{n\lambda's'}^3 
+c_{n\lambda s}^4 c_{n\lambda's'}^4) 
\nonumber\\
&+v_z (c_{n\lambda s}^4 c_{n\lambda's'}^1  
+c_{n\lambda s}^3 c_{n\lambda's'}^2
+c_{n\lambda s}^2 c_{n\lambda's'}^3  
+c_{n\lambda s}^1 c_{n\lambda's'}^4) 
\Big]\delta_{n,n'},
\label{Jz}
\end{align}
\end{widetext}
where $c_{n\lambda s}^{i=1,\cdots,4}$ are the coefficients of the wavefunction and the selection rules are given as $n\rightarrow n$.  When the optical field acts in the perpendicular $x$ or $y$ direction, $\boldsymbol E\perp\boldsymbol B$, we have
\begin{widetext}
\begin{align}
\langle\psi_{n\lambda s}|J_x|\psi_{n'\lambda's'}\rangle
=&\frac{e\sqrt 2\xi}{l_B} 
\Big[(-c_{n\lambda s}^1c_{n'\lambda's'}^1+c_{n\lambda s}^4c_{n'\lambda's'}^4) 
\sqrt{n+1} +(c_{n\lambda s}^2c_{n'\lambda's'}^2-c_{n\lambda s}^3c_{n'\lambda's'}^3) 
\sqrt{n}\Big] \delta_{n,n'-1} 
\nonumber\\
&+\frac{e\sqrt 2\xi}{l_B} 
\Big[(-c_{n\lambda s}^1c_{n'\lambda's'}^1+c_{n\lambda s}^4c_{n'\lambda's'}^4) 
\sqrt{n} +(c_{n\lambda s}^2c_{n'\lambda's'}^2
-c_{n\lambda s}^3c_{n'\lambda's'}^3) 
\sqrt{n-1}\Big] \delta_{n,n'+1}
\nonumber\\
&+ev(c_{n\lambda s}^1 c_{n'\lambda' s'}^2 -c_{n\lambda s}^4 c_{n'\lambda's'}^3)
\delta_{n,n'-1} 
+ev(c_{n\lambda s}^2 c_{n'\lambda's'}^1
-c_{n\lambda s}^3 c_{n'\lambda's'}^4)\delta_{n,n'+1}, 
\label{Jx}
\end{align}
\end{widetext}
and the selections rules are $n\rightarrow n\pm1$.  
We can see that (i) the selection rules are distinct for the different field configurations; (ii) for a specific configuration, the selection rules are the same for the linear bands and the inverted parabolic bands;  (iii) the selection rules have no restriction on the $s$ index.  Experimentally, the parallel configuration $\boldsymbol E\parallel\boldsymbol B$ corresponds to the Faraday configuration, while the perpendicular configuration $\boldsymbol E\perp\boldsymbol B$ is also feasible. 

In Fig.~\ref{Fig3}, we plot the optical conductivity Re$(\sigma_{\alpha\alpha})$ as a function of the photon frequency $\omega$ for the different linewidth $\tau^{-1}$ and the magnetic field $B$, with $\alpha=z$ in (a)$-$(c) and $\alpha=x$ in (d)$-$(f).  Similar to the DOS, to obtain $\sigma_{\alpha\alpha}$, the numerical integrations over $k_z$ are also needed.  Clearly, the peaks in Re$(\sigma_{\alpha\alpha})$ occur at the resonant frequency $\omega=\varepsilon_{n'\lambda's'}-\varepsilon_{n\lambda s}$, as labeled in Fig.~\ref{Fig3}.  In Re$(\sigma_{zz})$, the peaks of the transitions $n\lambda s\rightarrow n\lambda' \bar s$ and $n\lambda \bar s \rightarrow n\lambda's$ occur at the same $\omega$, with $\bar s=-s$, so they are twofold degenerate.  The twofold degenerate peaks also happen for the transitions $n\lambda s\rightarrow n+1,\lambda'\bar s$ and $n+1,\lambda \bar s\rightarrow n\lambda's$ in Re$(\sigma_{xx})$.  Moreover, when the polarization of the wavefunction $\psi_{n\lambda s}$ is taken into account, we find that the peaks of Re$(\sigma_{\alpha\alpha})$ are not spin-polarized, even at the $\Gamma$ point.

We consider the evolution of Re$(\sigma_{zz})$ with the magnetic field.  In both Figs.~\ref{Fig3}(a) and (b), the unambiguous peaks at the $\Gamma$ point are shown by the arrows.  
(i) When the magnetic field is weak [Fig.~\ref{Fig3}(a)], we can see that each LL transition owns two resonant peaks that are caused by the diverging DOS at the $\zeta_n$ and $\Gamma$ points.  As the $\zeta_n$ ($\Gamma$) point acts as the local energy minimum (maximum), the transition peak corresponds to the lower (upper) threshold of the transitions, together with a long tail due to the dispersive 3D LLs~\cite{Ashby}.  For $n\geq3$ LL transitions, the peaks at the $\Gamma$ point are close to the $\zeta_n$ point, so they are hardly to be distinguished.  More importantly, the peaks at the $\zeta_n$ point are stronger than those at the $\Gamma$ point, Re$(\sigma_{zz}^{\zeta_n})>$Re$(\sigma_{zz}^\Gamma)$. 
(ii) At intermediate magnetic field [Fig.~\ref{Fig3}(b)], for the LL transition $0\lambda\rightarrow 0\lambda'$, two resonant peaks that are similar to Fig.~\ref{Fig3}(a) can be seen, while for $n\geq1$ LL transitions $n\lambda s\rightarrow n\lambda's'$, there exist three resonant peaks due to the large Zeeman splitting $\Delta_{ns}$ (the middle peak is twofold degenerate). 
(iii) At strong magnetic field [Fig.~\ref{Fig3}(c)], for $n=0$ LL transition, only one resonant peak can be seen, as the $\zeta_n$ point has been merged with $\Gamma$.  For $n\geq1$ LL transitions, there exist the strong and weak peaks, where the former ones are related to the LL transitions with $s'=s$, while the latter ones are related to those with $s'=-s$.  The behavior of the strong and weak peaks reminds us of the previous magnetic-optic conductivity study in the line-nodal semimetal model~\cite{Y.X.Wang2017, R.Y.Chen2015b}, where the similar conclusions are reported.  We can also see that when $B$ increases, the first peak of the transition $0\lambda\rightarrow0\lambda'$ in Re$(\sigma_{zz})$ occurs at lower $\omega$ [Figs.~\ref{Fig3}(a)$-$(c)].  

Next we turn to Re$(\sigma_{xx})$. 
(i) When $B$ is weak [Fig.~\ref{Fig3}(d)], there are also two resonant peaks occurring at the saddle points in each LL transitions.  But the two peaks have comparable heights, Re$(\sigma_{xx}^{\zeta_n})\simeq$Re$(\sigma_{xx}^\Gamma)$, and hence are different from those in Re$(\sigma_{zz})$. 
(ii) At intermediate $B$ [Fig.~\ref{Fig3}(e)], four resonant peaks appear in the transitions  $0\lambda\rightarrow1\lambda's'$, which are caused by the existence of the two saddle points in the $n=0$ LL as well as the Zeeman splitting of $n=1$ LLs.  Note that these peaks also have comparable heights. 
(iii) At strong $B$ [Fig.~\ref{Fig3}(f)], two resonant peaks can be found in the LL transitions  $0\lambda\rightarrow1\lambda's'$, which are due to the further splitted $s'=\pm1$ LL branches.  The higher LL transitions occur at much larger photon frequencies and are not considered in Figs.~\ref{Fig3}(e) and (f).  In addition, when $B$ increases, the first peak of the transition $0\lambda\rightarrow1\lambda's'$ in Re$(\sigma_{xx})$ occurs at larger $\omega$ [Figs.~\ref{Fig3}(d)$-$(f)], which is different from Re$(\sigma_{zz})$.  

We analyze the effect of the impurity scattering that is phenomenologically represented by $\tau^{-1}$ on Re$(\sigma_{\alpha\alpha})$.  We can see that when $\tau^{-1}$ increases, the resonant peaks of Re$(\sigma_{\alpha\alpha})$ are blurred out and may be smoothened, which are consistent with the previous studies~\cite{Ashby, Y.X.Wang2020, W.Duan2019}.  When $\tau^{-1}\leq0.2$ meV, the double-peak structures at the saddle points are clearly seen in Re$(\sigma_{\alpha\alpha})$.  However, when $\tau^{-1}$ increases to 1 meV, in Re$(\sigma_{zz})$, the weak peaks at the $\Gamma$ point are completely wiped out and the strong peaks at the $\zeta_n$ point are retained  [Figs.~\ref{Fig3}(a) and (b)], while in Re$(\sigma_{xx})$, the two peaks will be merged into one [Figs.~\ref{Fig3} (d) and (e)].  Thus the double-peak structures at the saddle points will become single-peak ones in Re($\sigma_{zz}$) as well as Re$(\sigma_{xx})$.  Besides the impurity scattering, the increasing temperature can also smoothen the resonant peaks of Re$(\sigma_{\alpha\alpha})$~\cite{Ashby}.  So to capture the magneto-optical signatures of the saddle points, a clean ZrTe$_5$ sample and a low-temperature condition are required in the experiment. 

We point out that the qualitative relation of the peak difference at $B=0.4$ T [Fig.~\ref{Fig3}(a)], Re$(\sigma_{zz}^{\zeta_n})>$Re$(\sigma_{zz}^\Gamma)$, is consistent with the observations in the experiment~\cite{Y.Jiang2020}, where the $n\geq1$ LL transitions in the infrared region are focused on.  To explain their results, the authors used the rule-of-thumb formula 
\cite{X.Lu, M.Dressel}, where the matrix element of the current density operator is assumed to be a constant, 
\begin{align}
\text{Re}(\sigma)\sim\frac{\text{JDOS}}{\omega}. 
\label{rule}
\end{align}
The rule-of-thumb formula tells us that the peak shape is determined by the ratio between the joint DOS (JDOS) and the resonant frequency, where the inverse scattering rate has been included in the JDOS~\cite{X.Lu}.  They further assumed the different linewidth and chose $\tau^{-1}_{\zeta_n}=0.04\omega$ and $\tau^{-1}_\Gamma=0.06\omega$ to fit their experimental observations~\cite{Y.Jiang2020}, meaning that the electron experiences more scattering events at the $\Gamma$ point than the $\zeta_n$ point. 
Here through numerical calculations, we show that the qualitative relation can be well captured by using the full Kubo formula and assuming the same linewidth broadening $\tau^{-1}$ at the saddle points.  Actually, for the $n=0$ LL transitions, the matrix elements of $J_z$ at the saddle points are calculated as 
\begin{align}
&J_z^{0\lambda,0\lambda'}(\Gamma)=ev_z,
\\
&J_z^{0\lambda,0\lambda'}(\zeta_0)=\frac{e v_z (M-\frac{\xi}{l_B^2}+\xi_zk_{z0}^2) }
{\sqrt{(M-\frac{\xi}{l_B^2}-\xi_zk_{z0}^2)^2+v_z^2k_{z0}^2}}. 
\end{align}
Clearly we have $J_z^{0\lambda,0\lambda'}(\Gamma)<J_z^{0\lambda,0\lambda'}(\zeta_0)$ as  $(M-\frac{\xi}{l_B^2}-\xi_zk_{z0}^2)$ is comparable to $v_z k_{z0}$.  Combining the matrix element relation with $\frac{1}{\varepsilon_{0\lambda'}-\varepsilon_{0\lambda}}\big|_{k_z=0}<\frac{1}{\varepsilon_{0\lambda'}-\varepsilon_{0\lambda}}\big|_{k_z=k_{z0}}$, the peak difference is well understood.  Similar analysis can also be extended for other double-peak structures in the optical conductivity.  Thus the current density operator plays an important role in determining the peak shape of the optical conductivity and the rule-of-thumb formula may not be sufficient.  Based on these analysis, we suggest that at the saddle points, the peak difference in Re$(\sigma_{zz})$ and the almost equal peak heights in Re$(\sigma_{xx})$ root in the gapped Dirac semimetal model itself, which act as the intrinsic properties of the system.

\section{Discussions and Conclusions}

When the spin Zeeman effect is included in the system~\cite{Z.G.Chen, R.Y.Chen2015b}, $H_{sz}=-\frac{1}{2} g\mu_B B\sigma_z$, with $g$ being the Land\'e $g$-factor and $\mu_B$ the Bohr magneton, the total Zeeman splitting is strengthened. As a result, the critical magnetic field becomes $B_0^c=\frac{1}{e\xi+\frac{1}{2}g\mu_B}(M-\frac{v_z^2}{2\xi_z})$ and will be decreased.  For the optical conductivities, we have (i) all resonant peaks will move to lower photon frequency $\omega$ as the LL energies become lower, (ii) in the intermediate and strong magnetic field regions, the resonated peaks related to the $n>0$ LL transtions, such as the three peaks in the transitions $n\lambda s\rightarrow n\lambda's'$ of Re($\sigma_{zz}$), will become more widely separated, which favors the experimental observations.

To summarize, we have made a systematic study on the LLs and the optical conductivity Re$(\sigma_{\alpha\alpha})$ in the gapped Dirac semimetal model of ZrTe$_5$ under a magnetic field and reveal new physics.  The LL structures are determined by the interplay of the two quantities, $\Delta_n$ and $\Delta_{ns}$, and then the magnetic field can be separated into the weak, intermediate and strong magnetic field regions.  In the different magnetic field regions, the DOS and Re$(\sigma_{\alpha\alpha})$ exhibit distinct characteristics.  It is interesting to find that the strong magnetic field can drive the strong TI phase in ZrTe$_5$ enter the trivial insulator phase.  Among the characteristics of Re$(\sigma_{\alpha\alpha})$, two aspects are worth emphasizing: (i) the evolutions of the first set of peaks (related to the $n=0$ LL) in Re$(\sigma_{\alpha\alpha})$ with the magnetic field, and (ii) the peak difference of Re$(\sigma_{zz})$ at the saddle points in the weak and intermediate magnetic field, in sharp contrast to the almost equal peak heights of Re$(\sigma_{xx})$.   

In the experiment, compared with other factors, such as the temperature and interlayer distance, the magnitude of the magnetic field is more easily modulated and the strong magnetic field can be achieved by using the pulsed magnetic field.  So the magnetic field can provide an excellent probe in checking the dimensionality of the Dirac cone as well as the validity of the gapped Dirac semimetal model in ZrTe$_5$.  When the magnetic field acts in the plane~\cite{Y.Jiang2020}, the quantized LLs also exist, but cannot be solved analytically.  Instead, the LLs may be obtained numerically by discretizing the low-energy Hamiltonian on a lattice model~\cite{Y.X.Wang2019}.  We expect that the selection rules of $n\rightarrow n$ for the parallel optical field and magnetic field configuration $\boldsymbol E\parallel\boldsymbol B$ and of $n\rightarrow n\pm1$ for the perpendicular configuration $\boldsymbol E\perp\boldsymbol B$ are still valid, while the transport signatures for such field configurations are left as open questions that need more theoretical studies.

\section{Acknowledgments} 

We would like to thank Fuxiang Li for many helpful discussions.  This work was supported by the National Natural Science Foundation of China (Grant No. 11704157 and No. 11804122), and the China Postdoctoral Science Foundation (Grant No. 2021M690970).

\end{document}